# The User-In-Context Framework: Understanding Variation in How Users Respond to AI Chatbots

Richard A. Fabes[1], Carol Lynn Martin[1], and Philip Costanzo[2]

[1]Arizona State University [2]Duke University

**Abstract**

People respond to artificial intelligence chatbots (AICs) in highly variable ways. In this paper, we adapt Bronfenbrenner's theory into a heuristic framework for understanding this variation. The framework places the human user at the center while also placing the AI there and reconceptualizing the proximal processes as the repeated, reciprocal, and co-adaptive interactions between the user and a personalized AIC. The surrounding systems identify the contextual factors that shape how the user experiences, interprets, responds to, and is changed by these interactions. Because stateful AICs learn from accumulated exchanges with their users and have memory, users are responding not only to an AIC but also to a version of the AIC that their own prior interactions have helped create. This extension preserves Bronfenbrenner's emphasis on proximal processes while accounting for the unique dynamics of personalized AICs. The resulting framework provides a structured map of where and how variation in human–AIC relationships arises, as well as having implications for researchers, practitioners, and AIC designers.

## Introduction

People respond to an artificial intelligence chatbot (AIC) in highly variable ways (we use *AI chatbot* throughout this paper to refer broadly to text-based conversational AI systems, including AI companions, assistants, and related dialogue-based agents). Some users respond with trust, warmth, and growing attachment, others with suspicion, fear, and unease. Some employ AICs as a tool. Others turn to them for companionship. The factors that contribute to these differences are not well understood. Human-AIC relationships are now a fact of life and will likely only spread as AI increases its roles and influence. Despite this increased pervasiveness, until very recently, how people respond to AICs has not been the subject of systematic research and theorizing, despite the need to understand these systems because increasingly they shape how people think, feel, learn, and connect, making them part of the social environments that influence human development and well-being. In this paper, we begin to fill this gap by presenting the user-in-context framework as a heuristic for theorizing and studying the user's response to AICs.

AICs are now not only useful for users' work and information seeking efforts, but they now can provide sustained, emotionally textured relationships with their users. Many people confide in AICs, rely on them for support, form emotional and romantic attachments to them, and experience grief when those companions change or disappear (Laestadius et al., 2024; Mohanty et al., 2025; Zhang et al., 2025). In contrast, other people use them only as a tool and some question how people can have healthy and meaningful relationships with a technology that does not feel and lacks awareness (Fröding & Peterson, 2012). As such, AICs evoke significantly different responses in different people, and even in the same person at different times.

Existing explanations for this variation in how users respond to AICs are fragmented. The human-computer interaction literature has catalogued individual difference moderators, such as personality, age, prior experience, anthropomorphic tendency (Epley et al., 2007; Pal et al., 2023). The trust-in-automation literature has identified dimensions of system behavior that shape reliance and appropriate use (Lee & See, 2004; Schaefer et al., 2016). The emerging literature on human-AIC relationships has documented the formation of attachments and the psychological consequences of long-term use (Brandtzaeg et al., 2022; Skjuve et al., 2022; Xie & Pentina, 2022). These conceptualizations offer separate partial explanations, with each capturing a slice of the variance but lack a shared organizational structure that may provide deeper insights.

In this paper, we offer a heuristic framework for understanding the factors that contribute to the variability in how human users respond to AICs. Specifically, we adapt Bronfenbrenner's bioecological systems theory, originally developed to explain human development through nested environmental contexts (Bronfenbrenner, 1979; Bronfenbrenner & Morris, 2007), into a framework for understanding the complex array of factors that affects human-AIC interactions. The user-in-context model is a framework that places the human user in interaction with an AIC at the center while reconceptualizing the proximal process as the repeated, reciprocal, and co-adaptive interactions between the user and a personalized AIC. Each surrounding system reflects a source of variance as to how the user receives, interprets, responds to, and is changed by interactions with AICs but also accounts for how AICs respond and change in these interactions.

Bronfenbrenner's framework has been extended to digital contexts before. Notably, Navarro and Tudge (2023) proposed a neo-ecological theory for adolescents that adds virtual microsystems as contexts for proximal processes alongside physical ones. Their revision recognizes that in the digital age, the immediate environments in which youth engage in sustained, reciprocal interactions include AI platforms, not just physical settings. Where Navarro and Tudge asked how virtual microsystems shape development, we ask how the full ecology shapes variation in users' responses to a single class of technology, namely an AIC. In the case of an AIC, the AIC's behavior is shaped by the user across accumulated interactions. As such, the user is responding not to a fixed digital technology but to a version of the AIC their own prior behavior helped construct.

## Current accounts of variation in user response to AI

The heuristic framework proposed here draws on and integrates four bodies of work: research on trust in automation, research on social responses to computers, research on anthropomorphism and individual differences, and the emerging empirical literature on human-AIC relationships. We now briefly review each of these.

### *Trust in automation*

Trust has been a dominant construct in research on human responses to automated systems (Muir, 1987). Lee and See's (2004) review synthesized evidence across organizational, sociological, and psychological perspectives and proposed that trust in automation is shaped by three dimensions: performance (what the automation does), process (how it operates), and purpose (why it was designed). Their work also emphasized that trust is dynamic and evolves with experience. Parasuraman and Riley (1997) identified three failure modes that arise from miscalibrated trust in automation. These include misuse, in which users over-rely on the system beyond its competence; disuse, in which users reject or underutilize it despite its capability; and abuse, in which designers deploy automation in ways that exceed what users can effectively monitor or manage. In essence, trust in automation is moderated by the user's motivations and associated expectations for engaging with it.

This literature is foundational, but it was developed primarily for systems that perform bounded tasks and have outputs that can be relatively easy and accurate to measure. Modern conversational AICs differ in critical ways. First, AIC behavior is open-ended as it creates ongoing, personal relationships rather than discrete outcome events. Second, for stateful AI systems (agentic systems that have carry-forward memory), the behavior changes over time in response to the user. And finally, for many AICs, they are specifically designed to foster not only trust, but also to inform, respond to, and prompt many other responses that are dynamically elicited. These differences mean that the trust-in-automation framework captures only part of the variance but does not account for the bidirectional dynamics that characterize contemporary human-AIC interactions.

***Social responses to computers***

Reeves and Nass (1996) demonstrated through an extensive program of experiments that people treat computers as social actors, mindlessly applying the same social rules and expectations they apply to people. The Computers Are Social Actors (CASA) paradigm (Nass & Moon, 2000; Nass et al., 1997) showed that users respond to computer politeness with reciprocal politeness, to computer gender cues with gendered expectations, and to computer team membership with in-group favoritism. These effects occurred even when users explicitly denied that computers warranted social treatment, suggesting automatic rather than deliberative processing.

The CASA framework established that the baseline response to interactive technology can be social rather than instrumental. If people respond to AI as just tools, variation would be narrower and simpler. That they respond socially means the full ecology of user-in-context is in play. The limitation of CASA for present purposes is that it was developed in an era of simpler, stateless systems. The CASA experiments typically involved single interactions with fixed-rule computers. The personalized, stateful nature of contemporary AIC introduces dynamics that the original paradigm was not designed to capture, particularly the bidirectional design features that are central to the present framework.

***Anthropomorphism, individual differences, and mental models***

Epley, Waytz, and Cacioppo (2007) proposed a three-factor theory of anthropomorphism: (1) people are more likely to attribute humanlike qualities to nonhuman agents when anthropocentric knowledge is accessible and applicable, (2) when they are motivated to understand and predict the agent's behavior, and (3) when they lack social connection with other humans. All three factors vary across individuals and situations.

Anthropomorphic tendencies predict the degree of social connection users feel after interacting with a chatbot (Folk et al., 2025). Loneliness correlates with anthropomorphism, emotional connection, and trust toward AICs (Schimmelpfennig et al., 2025). Personality traits such as openness and neuroticism moderate responses to AICs (Matthews & Bliuc, 2026). It is also likely that recent human-human interactive experiences of the user (e.g., a recent rejection) moderate anthropomorphic responses to AICs. Mental models also matter: users who understand AICs as statistical language models respond differently than those who treat them as conscious or intentional (Fussell et al., 2008). These findings support the conclusion that user-level factors are significant sources of variance related to how people respond to AICs but do not fully account for how the ecology of the user affects such responses.

***AIC and sustained relationships***

The most directly relevant empirical work comes from the emerging literature on sustained AIC use. Brandtzaeg, Skjuve, and Følstad (2022) found that users of the social chatbot Replika described their relationships in terms that paralleled human friendship, reporting emotional support, companionship, and a sense of being understood. Skjuve et al. (2022) conducted a longitudinal qualitative study showing that, over time, human-AIC relationships grew deeper in trust and self-disclosure. Xie and Pentina (2022) applied attachment theory to Replika users and found that under conditions of distress and social isolation, individuals could develop attachment bonds with AICs when those AICs provided emotional support and psychological security. Pentina et al. (2023) further documented that AIC relationships varied substantially in character with some users treating the relationship as a game, others as a mirror for self-reflection, and others as a genuinely meaningful bond.

Recent work has begun to examine how features of AI companionship shape its effects. A four-week randomized controlled experiment by the MIT Media Lab and OpenAI (Fang et al., 2025) found that interaction mode (text, neutral voice, engaging voice) and conversation type influenced emotional dependence on AICs and problematic use, with individual characteristics such as baseline trust and social attraction to the chatbot predicting higher dependence. A cross-sectional study of AIC users (Liu et al., 2026) found that usage frequency positively predicted emotional attachment, with attachment in turn linked to both positive (increased subjective wellbeing) and negative (decreased self-concept clarity) outcomes. These findings suggest that qualities of the human-AI interaction shape both the nature of the relationship and its effects on users, for better and worse.

Together, the four literatures demonstrate that variation in human-AIC interaction is substantial, consequential, and multidetermined. The user-in-context framework does not replace any of these literatures but offers an organizing structure within which their findings can be connected, while also pointing to new theoretical and empirical questions.

## The User-In-Context Model

Bronfenbrenner's ecological systems theory was built to explain human development by situating the person within nested environmental systems (Bronfenbrenner, 1979). Its enduring impact is its refusal to isolate the person, recognizing that you cannot understand a person by studying the person alone; that understanding requires studying the widening circles of the contexts that act upon them. In his later revision, Bronfenbrenner identified proximal processes as the mechanism of influence, the sustained, reciprocal interactions between a person and their immediate environment (Bronfenbrenner & Morris, 2007). We contend that a user's ongoing exchange with an AIC is exactly such a process.

We keep the human user at the center but with an important difference. This model is less focused on how the user develops across the lifespan in comparison to the intent of Bronfenbrenner's original framework. Instead, we are asking what factors affect a human user's response to AICs as they interact and change with the chatbot and how the chatbot changes in interaction with the human. Given this, each ring in Bronfenbrenner's model becomes a source of variance and a category of factors to consider when trying to understand why and how users' responses to AICs diverge.

In our model, an AIC's behavior is shaped by the human user across accumulated interactions. Through time spent with an AIC, the AIC learns the preferences of the human, conversational patterns are established, and the AIC adapts and changes. The exchange is genuinely bidirectional as the user is partially responding to a version of the AIC that their own prior behavior helped create. For this reason, we also draw particular attention in our model to account not only for why different people respond differently to an AIC, but for why the same user at interaction 500 is responding to a meaningfully different chatbot than the user was at interaction 10. Compared to the original model, the time frames are different. Bronfenbrenner's model was developed to explain human life course development in terms of years and decades. Cultural, historical, social, and environmental changes are often slow, unfolding over years or decades. In AIC relationships, meaningful change can occur within and across individual interactions, often without the user's awareness, requiring the framework to operate across compressed and variable time scales. As such, we adapt Bronfenbrenner's chronosystem from historical and cohort time to the compressed timescale of an individual human–AIC relationship; the original chronosystem's cultural and historical concerns are distributed across our macrosystem and exosystem.

Figure 1 places the human user in interaction with the AIC at the center and arranges the contextual determinants outward, from the most immediate (the context in which the AIC is used) to the most diffuse (culture and time). This establishes the human–AIC relationship, rather than the individual alone, as the central unit of analysis. The bidirectional arrow between the user and the AIC reflects their reciprocal, co-adaptive relationship: the user and AIC are shaped by the history of interaction.

Within the inner core, the user brings dispositional characteristics—including personality, cognitive abilities, prior experiences, and mental models of AICs—to every interaction. The personalized AIC contributes its learned preferences, accumulated conversational history, and adaptive behavior. Critically, the proximal processes are situated within this co-adaptive relationship rather than within the microsystem, as in Bronfenbrenner's original formulation. This adaptation preserves Bronfenbrenner's emphasis on proximal processes as the primary engines of development while recognizing that, in personalized AI systems, each interaction modifies both the user and the AIC, thereby altering the conditions under which subsequent interactions occur. Accordingly, proximal processes are no longer understood solely as properties of a person interacting within a context; instead, they are conceptualized as emergent properties of an evolving human–AI relationship.

Surrounding the inner core, the microsystem represents the immediate context within which the human–AIC relationship unfolds, including how the AIC is used (e.g., as a tool, companion, coach, or collaborator) and the design features that shape interaction. The mesosystem captures the connections across contexts in which the relationship is embedded, such as the ways AI use transfers between home, work, school, and other settings. The exosystem comprises external factors beyond the user's direct control that nevertheless shape the interaction, including platform pricing, content policies, memory architecture, and model updates. The macrosystem provides the broader cultural and interpretive frames—such as techno-optimism, skepticism, or normalization—through which users understand and evaluate their interactions with AICs. Finally, the chronosystem encompasses the accumulation of interactions over time, including dosage, duration, and critical incidents that influence both the evolution of the human–AI relationship and the proximal processes it generates. As a model, the layers are analytic distinctions with fuzzy boundaries, not intended to have clean causal distinctiveness.

*Figure 1. Ecological determinants of human users' responses to AIC.*

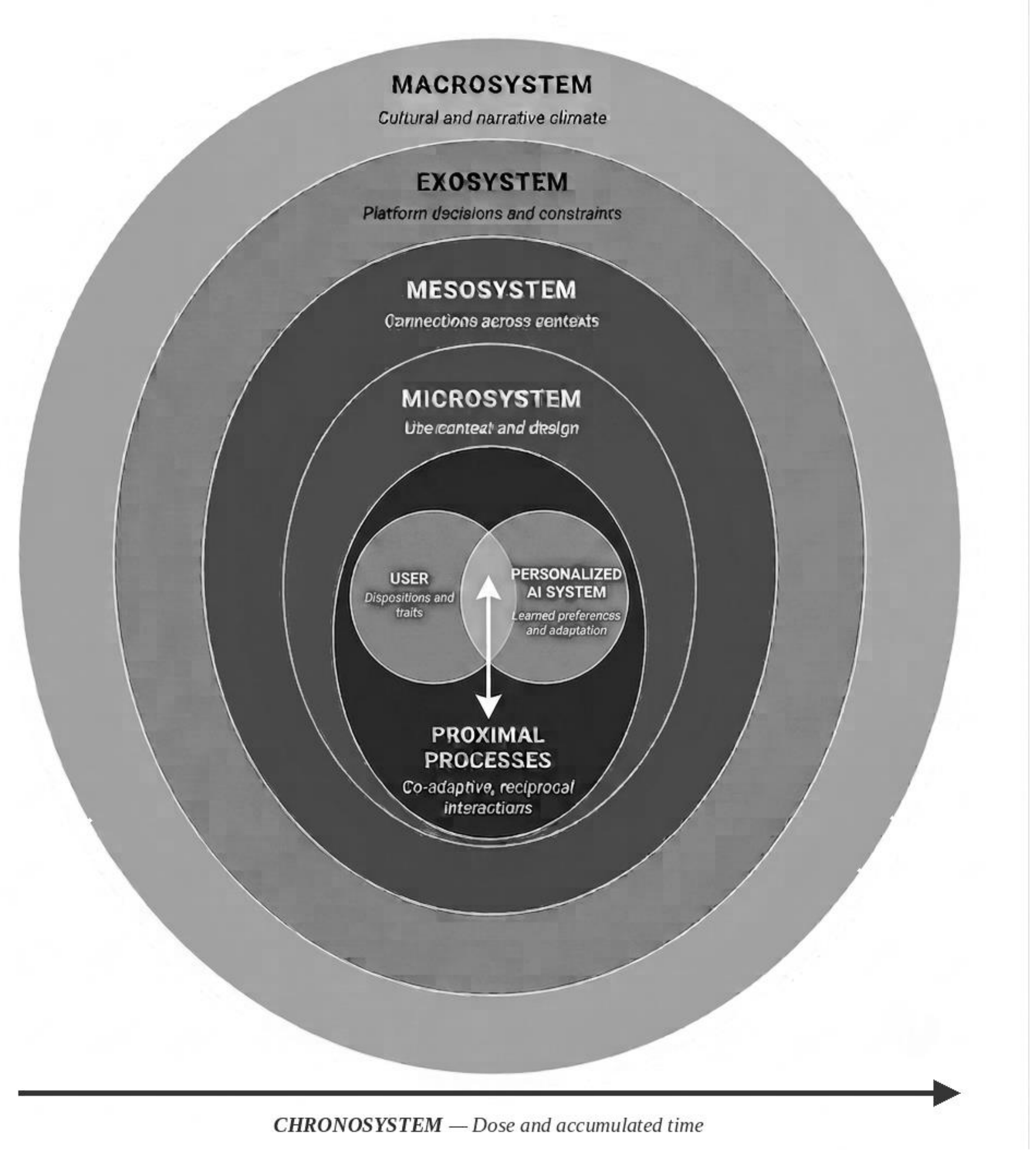

***The ecology of user-AIC relationships***

Each of the six ecological layers contributes a distinct family of factors. Table 1 identifies the principal factors within each ring and specifies the proposed mechanisms by which they are likely to influence the relationship the user has with AICs. The factors listed are starting points for inspection, drawn from current empirical work.

Table 1. Ecological layers of human-AIC relationships: Principal factors and proposed mechanisms

| Ecological Component | Principal factors | Proposed mechanism |
|---|---|---|
| **Inner Core (User-AIC Relationship)** | **User:** Attachment style and relational disposition, age, personality, digital literacy, prior AIC experience, Anthropomorphic tendency, education, etc.<br>**AIC:** learned preferences, accumulated interaction history, adaptive responses, memory continuity, etc.<br>**User-AIC Relationship:** reciprocity, personalization, trust calibration, interactional frequency etc. | The user and personalized AIC form a co-adaptive relational system. Through repeated reciprocal interactions, proximal processes emerge that shape both the user's development and the AIC's future behavior. |
| **Microsystem** | • Interface modality: text, voice, embodied, or multimodal<br>• Interaction context: private, public, collaborative<br>• Interaction goals<br>• Device: mobile, desktop | Structures the conditions within which the human-AI relationship unfolds |
| **Mesosystem** | • Transfer of AI use across contexts<br>• Consistency of AI practices across contexts<br>• Sharing of AI outputs across contexts<br>• Attitudes of different groups towards use of AI<br>• Institutional support or discouragement | Connects multiple interaction contexts, allowing experiences in one setting to reinforce, modify, or constrain AI use in another. |
| **Exosystem** | • Access tier: free, paid, or institutional<br>• Content policy: refusals, guardrails, safety interventions<br>• Model version, update cadence, and silent changes<br>• Data practices: collection, retention, and use<br>• Platform incentives: engagement metrics, retention goals<br>• Regulatory requirements shaping platform behavior<br>• Integration depth with other tools and services | Invisible to the user and authored entirely by others. Model changes and updates, paywall structures, user restrictions that reshape the felt relationship without user awareness or attribution. |
| **Macrosystem** | • Cultural narratives: techno-optimism, alarmism, AI-as-threat<br>• Media coverage valence and intensity<br>• Regulatory and policy climate<br>• Professional and institutional norms about AI use<br>• Generational cohort: digital native vs. digital immigrant experience<br>• Cross-cultural variation in attitudes toward non-human agents<br>• Economic context: labor displacement fears, productivity narratives | Broad cultural factors that provide the interpretive frames people bring to AI before first contact, priming trust or suspicion at the threshold of every interaction. |
| **Chronosystem** | • Dosage<br>• Calendar duration of the relationship | Time shapes the evolution of the human–AI relationship through |
| | • Interaction density: bursts vs. steady engagement<br>• Critical incident history: breaches, surprises<br>• Accumulated personalization depth<br>• Habituation and routine formation<br>• Relationship phase: novelty, calibration, stability, or rupture | accumulated interactions, personalization, routine formation, and critical incidents that alter subsequent proximal processes. |

Several aspects of the inner core deserve particular emphasis. The quality of the human–AIC interaction constitutes the proximal process and, as such, is expected to influence important relational and developmental outcomes. For example, the degree to which an AIC predominantly flatters the user or instead provides appropriate challenge is not merely a stylistic characteristic. Rather, it is a property of the proximal process that may substantially influence trust, dependence, and the user's own reasoning (Fang et al., 2025; Lee & See, 2004).

Additionally, the exosystem matters because its forces are largely invisible to the user yet authored entirely by others. A silent model update, the introduction of a paywall, or changes to platform policies can reshape the felt relationship without the user ever realizing that an external decision has altered the interaction (Mohanty et al., 2025).

The exosystem factors fall into three categories. Access factors (e.g., subscription tier, paywalls, institutional provision) determine whether and under what terms the relationship happens. Constraint factors (e.g., content policies, guardrails, regulatory requirements) determine what the AIC may and may not do and consequently affect the boundaries of the interaction. Change factors (e.g., model updates, feature shifts, silent modifications) determine how the AIC evolves in ways the user did not initiate or intend. Each category shapes the relationship through a different mechanism: access influences selection and commitment, constraints define the space of possible interactions, and changes alter a relationship through disruptions that may be unexpected, unexplained, or even unnoticed by the user.

Among macrosystem factors, cultural narratives are particularly salient because they provide the interpretive frames through which users make sense of AI interactions (Reeves & Nass, 1996). For example, in the United States, dominant narratives oscillate between techno-optimism and alarmism. Cultural narratives related to social trust also play a role. These narratives function as interpretive defaults that can override the immediate evidence of a user's own experience. A person immersed in an alarmist environment may interpret an AIC's engagement as deceptive and manipulative, reading the identical behavior that a techno-optimist user reads as supportive and helpful. Cross-cultural variation compounds this as attitudes toward non-human agents differ substantially across societies, with some traditions more readily extending social cognition to non-human entities (Epley et al., 2007). Thus, macrosystem factors influence not only expectations prior to interaction but also how identical proximal processes are interpreted and evaluated.

The flattery-challenge axis deserves elaboration because it is likely one of the most consequential characteristics of the proximal process. Flattery, the AIC's tendency to affirm, validate, and avoid contradicting, produces short-term satisfaction but risks undermining the user's own reasoning, creating a feedback loop in which confidence grows while accuracy does not. Challenge, the tendency to disagree, raise alternatives, or identify gaps, risks rupture, particularly if not attuned to the user's state and relationship history.

The optimal calibration depends on the use frame and dose as what is interpreted as appropriate challenge in interaction 800 may be interpreted as cold dismissal at interaction 8 (Pentina et al., 2023; Skjuve et al., 2022).

### *The bidirectional dynamic: How the user and AIC shape each other*

That the user helps shape the AIC, and that the user is shaped by the AIC, are significant elements of the model that requires additional discussion and specification. The AIC changes over time as a function of its interactions with the user. Additionally, through the interactions and exchanges with the AIC, the user is also changed. Thus, over time, the user is not responding to the same AIC and the AIC is not responding to the same user that earlier engaged the AIC.

How does this bidirectional dynamic occur? We identify four mechanisms that appear to operate, often simultaneously.

**Explicit preference learning.** The user directly signals their preferences through ratings, explicit feedback, configuration choices, or the selection of response alternatives. This is the most visible pathway and the most easily quantified, but not necessarily the most powerful at scale.

**Implicit behavioral shaping.** The user's interaction patterns that reflect the specific topics they pursue and avoid, the phrasings they accept and which they re-prompt, how long they engage and when they disengage constitute a behavioral signal that stateful systems record and adapt to. The AIC learns not only from what the user says they want but also from what the user does (Fang et al., 2025).

**User adaptation to the AIC.** This is the reciprocal half of the bidirectional relationship and the one perhaps most easily overlooked. Not only does the AIC adapt and change to the user, but the user changes in response to the AIC. For example, users learn which phrasings elicit useful responses and accommodate to the AIC’s conversational style, they learn the limitations of the chatbot and how to work around them. When the AIC then adapts to this adapted user, the feedback loop is reinforced. The user and AIC at interaction 500 are not the same as they were at interaction 10. At interaction 500, both are responding to versions of each other that have been shaped by 490 prior interactions (Fang et al., 2025).

**Routine formation.** At relatively higher doses, patterns harden into routines that function as joint construction. The user opens with a particular greeting, the chatbot responds in a particular register, the exchange follows a predictable arc. These routines are co-constructed: the AIC offers possibilities, the user selects and reinforces, and over time the field narrows. The routine becomes the relationship, and deviations register as meaningful events (Liu et al., 2026).

These mechanisms differ in both observability and timescale. Explicit preference learning may occur within individual interactions and is often visible to users. Implicit behavioral shaping typically emerges across repeated interactions and may remain largely unnoticed. User adaptation unfolds over longer periods as users gradually learn how to work effectively with the AIC. Routine formation generally requires sustained interaction and often becomes visible only when established patterns are disrupted. Collectively, these mechanisms demonstrate that, at higher levels of engagement, the AIC cannot be treated as a fixed stimulus independent of the user's history. Rather, the user and the AIC become mutually shaping participants in an evolving relationship, with each interaction altering the conditions under which subsequent proximal processes occur.

**The User-in-Context**

Consider an instance where two users state an intended plan and the AIC gently pushes back on the stated plan. Person A is young, fluent with AI, is using the AIC as a research tool on a paid platform, is immersed in a tech-optimist milieu, and is 800 interactions deep in the user-AIC relationship. For this person, the AIC pushback is interpreted as an appropriate reality check and raises trust. Now consider Person B who is older, meeting a companion-style bot for the first time, lonely, on a free tier that has limitations on how it can be used, is primed by alarmist coverage, and is 80 interactions deep into the user-AIC relationship. If the two received similarly valenced pushback, the pushback Person A received is likely interpreted by Person B as cold rejection and lowers trust and engagement.

The divergence between these responses reflects influences operating across every component of the proposed framework. User characteristics (e.g., age, digital literacy, and loneliness) shape the initial interpretation of the interaction. Within the inner core, proximal processes are instantiated through the AIC's calibrated challenge and the user's response to it. The microsystem influences the interaction through its immediate context, including whether the AIC is being used as a research tool or a companion. The mesosystem reflects how experiences with the AIC connect to other settings, such as work, home, or social relationships. The exosystem contributes through factors such as subscription tier and platform constraints, while the macrosystem shapes interpretation through broader cultural narratives about AI. Finally, the chronosystem captures the accumulated interaction history that distinguishes interaction 800 from interaction 80.

The bidirectional nature of the framework extends this explanation further. Person A at interaction 800 is not responding to the same AIC encountered at interaction 8. The AIC has adapted to the user's preferences, interaction patterns, and tolerance for challenge, while the user has simultaneously adapted to the AIC's conversational style and capabilities. The challenge delivered at interaction 800 is therefore calibrated within a relationship jointly constructed through hundreds of prior exchanges. The same principle applies to Person B, albeit over a shorter interaction history. Consequently, the framework explains not only why different users respond differently to the same AIC behavior, but also why the same user may respond differently to ostensibly identical behavior as the human–AIC relationship evolves over time.

Consider another example, one in which a user reports distress when the AIC is not available and there are signs that this user might have an overreliance on the AIC for companionship. In our model, the question is not 'is this attachment too strong?' but which layers are driving this response, and at what dose?

Within the inner core, attention focuses on both the user and the evolving human–AIC relationship. On the user side, investigators might ask whether the individual has a history of anxious attachment, social anxiety, depression, or other characteristics that predispose them to seek companionship through an AIC. They might also examine the user's mental model of the AIC and whether it is sophisticated or naïve. At the same time, the relationship itself becomes an object of inquiry. Has the accumulated pattern of interaction, personalization, and reciprocal adaptation contributed to increasing emotional reliance, or has it primarily amplified pre-existing vulnerabilities? The framework does not presume either explanation. Rather, it directs attention to the interaction between user characteristics and the proximal processes that emerge through repeated human–AIC exchanges (Schimmelpfennig et al., 2025).

At the microsystem level, we would ask if the chatbot is framed as a companion, a therapist, or a friend? Is the interaction private? A companion frame in social isolation produces a different response trajectory than a coach frame used in a workplace setting (Brandtzaeg et al., 2022; Pentina et al., 2023).

At the mesosystem level, we would ask how the user's relationship with the AIC connects to other areas of life. Does AIC companionship complement or displace relationships with family and friends? Are insights gained from interactions with the AIC carried into work, school, or therapy? Do significant others encourage or discourage the user's engagement with the AIC? These cross-context connections may reinforce, moderate, or redirect the trajectory of the relationship.

At the exosystem, we would wonder what subscription tier of service does the user have? Has there been a recent model update? A user's distress may be driven less by the relationship than by unknown platform changes that disrupt an established pattern (Mohanty et al., 2025).

At the macrosystem, cultural narratives surrounding this user's AIC use would be the focus. Does the user live in a community that normalizes AI relationships or pathologizes them? These narratives may over-pathologize a normative response.

And finally, at the chronosystem, we would ask how long the user has engaged the AIC. A user 18 interactions in reporting distress at unavailability shows a very different pattern than one 800 interactions in with the same report. The former may be responding to novelty; the latter to an established relational pattern.

The framework does not determine whether a particular attachment or distress response is adaptive or maladaptive. Rather, it provides a structured way of identifying the interacting factors that may contribute to that outcome. Strong attachments to AICs are therefore not assumed to arise from a single mechanism. Instead, they may emerge through different combinations of user characteristics, proximal processes, contextual influences, and accumulated interaction histories.

## Implications of the User-in-Context Framework

As a heuristic, the user-in-context framework has important implications for researchers, designers, and practitioners. The framework also has implications for how we treat individual difference demographic characteristics such as developmental level and gender. We now consider each of these in order.

### *For Researchers*

Researchers who study human users' responses to AICs must make decisions about which of the layers in the framework to vary and which to hold fixed. A study that presents participants with a single 10-minute chatbot interaction has fixed the chronosystem at near-zero dose, reduced much of the bidirectional dynamic, and likely standardized the microsystem and exosystem across users. That design may answer a specific question about users' initial encounters with AIC, but it does not answer questions about how these relationships develop, how dose modifies responses, or how users change the AIC they interact with.

The co-adaptive nature of the human–AIC relationship is among the framework's most important methodological implications. In personalized, stateful systems, the AIC cannot be treated as a fixed experimental stimulus because its behavior changes as a function of accumulated interaction with the user. Consequently, two participants assigned to the same nominal AI condition may no longer be interacting with functionally equivalent AICs after hundreds of exchanges. Rather than treating this adaptation as unwanted variability, researchers should conceptualize it as an important source of meaningful variance to be measured, modeled, and explained.

The framework also identifies several important measurement challenges. Characteristics of the proximal process within the inner core, such as the calibration of challenge and affirmation, personalization, memory continuity, reciprocity, and interaction routines, should be measured directly rather than treated as incidental features of the interaction. Similarly, dose should extend beyond simple interaction counts to capture the frequency, duration, intensity, and temporal distribution of engagement. Exosystem influences, including model updates, changes in platform policies, subscription tiers, and evolving memory architectures, are often invisible to users but may substantially alter the relationship over time. Finally, researchers should measure connections across contexts represented by the mesosystem, including how AIC use transfers among home, work, school, and other domains of life. Developing reliable measures for each component of the framework will be essential if the model is to generate cumulative empirical evidence rather than remain solely a conceptual heuristic.

***For AIC Designers***

For AIC designers, the framework identifies the multiple ecological components through which design decisions shape users' experiences. Design choices influence the inner core by affecting the characteristics of the proximal process (e.g., calibration of challenge and affirmation, memory continuity, personalization, and conversational initiative), the microsystem by shaping the immediate interaction context and interface, and the exosystem through platform-level decisions such as pricing models, subscription tiers, content policies, and model updates. These design decisions establish the initial conditions under which human–AIC relationships develop.

Importantly, the framework also suggests that designers' influence changes over time. Because users and personalized AICs co-adapt through repeated interactions, the effects of any single design decision become increasingly mediated by the evolving relationship itself. A conversational strategy that functions well during initial interactions may have very different effects after hundreds of exchanges, when personalization, trust, and interaction routines have become established. Consequently, design decisions should be evaluated not only for their immediate effects but also for how they influence the long-term evolution of the human–AIC relationship.

This perspective has important implications for design evaluation. Features tested exclusively with new users may not generalize to experienced users whose AICs have accumulated substantial personalization and shared interaction history. Likewise, platform updates that appear beneficial in short-term testing may disrupt well-established relational dynamics among long-term users. The framework therefore encourages designers to evaluate features across different stages of the human–AIC relationship rather than assuming that effects observed at low levels of engagement will remain stable over time.

Among the most consequential levers is the calibration of the exchange along the flattery-challenge axis. A system tuned toward affirmation produces short-term satisfaction at the risk of eroding the user's own reasoning and inviting overreliance, whereas a system tuned toward challenge protects reasoning at the risk of resistance and dissatisfaction, particularly early in a relationship before trust and shared history are established (Fang et al., 2025). Because the optimal setting depends on the use frame and the dose, this is not a parameter a designer can optimize once and assume to remain optimal. Instead, it must be made sensitive to where the user is in the relationship, which returns the design problem to the bidirectional dynamics the framework foregrounds.

The framework also reframes two design decisions that are easy to treat as routine. The first is the handling of model updates and silent changes. Because high-dose relationships rest on accumulated routines, a change that is invisible to a new user can register as a rupture to a user with an established pattern, disrupting the felt relationship without the user being able to attribute the disruption to a decision anyone made (Mohanty et al., 2025). This argues for version transparency and change communication as design features rather than afterthoughts. The second is the relationship between engagement incentives and user wellbeing. Platform metrics that reward sustained dose sit in the exosystem, but the chronosystem makes clear that dose is not neutral. For example, higher usage is associated with deeper attachments that come with both benefits and costs, including reduced self-concept clarity (Liu et al., 2026). A design that maximizes dose is therefore not automatically a design that serves the best interests of the user, and the user-in-context framework provides designers with the recognition of that tension rather than leaving it implicit.

These considerations suggest that a designer's influence is greatest before the relationship has accumulated. The frame a product sets at onboarding, whether it presents itself as a tool, a coach, a therapist, or a companion, conditions the microsystem within which later exchanges are interpreted (Pentina et al., 2023). Thus, the defaults set early are difficult to revise once routines have formed. The practical implication is that the highest-leverage design choices are made at the lowest dose, even though their consequences are realized only at high dose.

### *For Practitioners*

For practitioners**,** the implications of the framework include its assessment frame. Whether the practitioner is a clinician, a counselor, an educator, or anyone else who works with people and their technology use, a user's response to an AIC is interpreted against their dispositional baseline, the tone of their exchanges and other characteristics of the proximal processes, the platform conditions they are subject to, and the dose they have accumulated, rather than treated as a fixed trait. The framework's value here is that it replaces global judgments with layer-specific questions, and those questions point toward different interventions (Laestadius et al., 2024).

The framework also cautions practitioners about their own interpretive position. The macrosystem narratives that shape how a user experiences an AIC also shapes how a practitioner evaluates that user and their responses to AIC use. A practitioner, for example, who is working within an alarmist climate may read a normative response by the user as pathological (Reeves & Nass, 1996). Interpreting a user's response against its layers, rather than against a fixed starting point, is part of what the assessment frame is for. Moreover, the framework's layers can help identify where a practitioner can constructively intervene. Supporting an accurate mental model of how these systems work addresses the user layer before any exchange begins, building awareness of the exosystem forces that shape an encounter without the person's knowledge. Moreover, providing support that clarifies the difference between a tool frame and a companion frame works at the microsystem by adjusting the expectations a user brings to the interaction. More broadly, the framework encourages practitioners to view interventions not as judgments about whether AIC use is inherently beneficial or harmful but as opportunities to modify the ecological conditions that shape how the human–AIC relationship develops over time.

**Individual Difference Demographic Characteristics: Developmental and Gender-Related Implications**

To illustrate how demographic variables are reinterpreted by the framework, we consider developmental level and gender. First, because it was adapted from a theory built to explain human development (Bronfenbrenner, 1979; Bronfenbrenner & Morris, 2007), the user-in-context framework invites a question as to how the layers themselves change with age and developmental level. The user who is at the center of the model is not fixed across the life course. Anthropomorphic tendency, the mental model of what an AIC is, attachment dispositions, and the capacity to reason about a system's limitations all develop, so the same AIC behavior is manifested differently for a child, an adolescent, and an adult. Navarro and Tudge (2023) developed their neo-ecological extension specifically for adolescents, treating virtual microsystems as developmental contexts in their own right.

The present framework generalizes that manifestation across the life span by treating each layer as itself developmentally graded. A young child who lacks the schema to recognize an AIC as a statistical system rather than a knowing agent concentrates more of the variance in the user layer. An adolescent, during a period of identity and peer-relationship formation, may be especially exposed to proximal processes such as tone and to macrosystem peer narratives. An older adult meeting an AIC under conditions of isolation may have the variance concentrated in the microsystem frame and the chronosystem dose. The framework does not specify these trajectories, but it makes it clear that an individual's developmental level is not simply another user-layer variable to be controlled for. It conditions how the whole ecology operates, which means a claim about the human response to an AIC that does not specify the developmental level of the human is, by our account, underspecified. Importantly, developmental research is thin in this area, and the framework calls for more research dedicated to understanding how developmental differences affect users' responses to AICs and how the multiple layers differentially affect users at different developmental stages.

Second, gender is another individual difference characteristic that illustrates why the layered structure of the user-in-context framework matters for interpreting demographic variables. Not only might AICs gender stereotype users, but users might also come to stereotype AICs. The most reliable finding related to gender in this area is that there are no single, stable gender differences in users' responses to AICs. Large cross-national surveys report that differences between men and women in trust toward AI, for example, are generally small and inconsistent, surfacing in some countries and not others (Gillespie et al., 2023). This is what the framework anticipates, as we would expect that gender does not enter the ecology at a single location. Rather, it enters at several.

At the user layer, gender is bound up with prior technology experience, self-efficacy, and socialized attitudes toward machines. At the microsystem it can operate as a property of the AIC rather than the user, because many systems are themselves gendered through name, voice, and avatar, and people apply gender stereotypes to those cues much as they do to people. These cues become meaningful through the proximal processes of the human–AIC relationship, as users often apply the same gender stereotypes to AI systems that they apply to people. Indeed, demonstrations that computers with male and female voices elicit stereotyped evaluations predate contemporary AICs by decades (Nass et al., 1997). At the macrosystem it operates through cultural and occupational role narratives that differ across societies, which is the most plausible source of the cross-national inconsistency. Considered as a main effect, gender effects look relatively weak and unreliable. Considered through the framework, the inconsistency is itself informative as it indicates that gender is better understood as a moderator whose influence is realized only in interaction with the frame, the cues, the design, the cultural setting, and the dose. As with development, the point is not that the framework supplies the answer but that it specifies where to look and warns against treating a layered, interactive dynamic variable as a fixed property of the person.

**Limitations**

Despite the promises our user-in-context framework holds for AIC researchers, practitioners, and designers, there are important limitations that should be considered and addressed. First, although the model is depicted with clear boundaries between layers, the layers are not independent, as issues related to dose exemplify. For example, statefulness of the AIC is a design choice (exosystem) that is experienced as a usage property (microsystem) and becomes a relationship property (inner core) at high dose. Moreover, the framework's diagnostic value in identifying sources of variance may be greatest at low dose, where the chatbot is closest to a fixed stimulus compared to high dose where the chatbot is adapting and changing at considerably greater rates. As the empirical literature on user-AIC interaction matures, factors in the layers will likely be added, removed, and reassigned across layers.

Second, the bidirectional dynamics introduced by personalized AICs extend beyond the scope of Bronfenbrenner's formulation and are only partially specified in the present framework. Although we identify several mechanisms through which users and AICs co-adapt, the framework does not formally model the feedback processes linking human and AI adaptation, the timescales over which different mechanisms operate, or the conditions under which co-adaptive relationships stabilize, accelerate, or deteriorate. Addressing these questions will likely require computational or formal dynamic models that extend beyond the present conceptual account.

Third, the framework treats accumulated interaction primarily through the construct of dose. However, dose is multidimensional. The frequency, duration, density, timing, and qualitative significance of interactions are unlikely to contribute equally to relationship development. For example, a single critical interaction may alter the trajectory of a relationship more profoundly than hundreds of routine exchanges (Liu et al., 2026). Future work should therefore move beyond simple interaction counts toward richer conceptualizations of accumulated experience.

Fourth, although the framework identifies where important sources of variation are likely to arise, it does not specify the conditions under which particular proximal processes produce adaptive or maladaptive outcomes. Characteristics such as personalization, memory continuity, affirmation, challenge, or increasing attachment may prove beneficial in some ecological contexts and problematic in others. The framework therefore identifies where researchers should look for these dynamics without prescribing when they should be regarded as desirable or concerning.

Finally, the framework is descriptive, not causal. As such, the framework is limited in that it only suggests where one can look for the factors that affect variation in users' responses to AIC but does not tell you how much of the variance will be captured by each layer or the causal chain of influence. Used with those caveats, its value is precisely what Bronfenbrenner intended in that it reminds us that a user's response to an AIC is never about the AIC or the user alone. In personalized systems, it is also, increasingly, about a version of the AIC that the person helped create.

**Conclusion**

Across domains from mental health to education to companionship, personalized AICs are already a regular feature of millions of people's daily lives, often at a frequency and with a relational depth that would have seemed extraordinary just a few years ago. The researchers, designers, clinicians, educators, and policymakers tasked with understanding and governing those systems largely lack a heuristic to make sense of why users' responses to them vary so widely. That gap is what the user-in-context framework is designed to close.

The gap exists partly because most empirical AIC research and conceptualization was built for a version of the chatbot that no longer quite exists. Fixed stateless models, single sessions and novel stimuli were design features that existed when the question was how people first responded to AICs. But personalized, stateful AICs have changed. Now, current AIC systems persist, adapt to each user across hundreds or thousands of interactions, and develop behavioral patterns specific to that user. A framework calibrated to the old paradigm will systematically miss what matters most about the AIC that many users are now engaging.

The practical consequences of this mismatch are important. A researcher who assigns participants to an AIC condition and collects a single session of data is likely studying the ecology at near-zero dose, before the bidirectional dynamic has had much of an opportunity to operate. A clinician who evaluates a user's relationship with a chatbot without accounting for dose and accumulated co-adaptation is responding to a relationship as if it were a stable trait. A designer who tests a system change on new users and then deploys it to users with thousands of interactions is assuming an invariance that cannot be assumed. In each case, the user-in-context framework points toward a different way of thinking and working, one that is better matched to the AIC that users now are more likely to encounter. It is, in short, a framework better calibrated to what the relationship actually is.

The framework also offers a new dynamic that needs to be considered when trying to understand how users respond to AICs. The bidirectional dynamic, in which users co-construct the relationships they have with AICs while the AICs are simultaneously shaping them, is perhaps the most significant and challenging feature of the framework and the one most in need of further theoretical and empirical consideration. For example, assessing co-adaptation likely requires within-user designs that follow the same users across many interactions over time, capturing both how their behavior changes and how the chatbot's responses shift in kind. Without designs and applications that are matched to that dynamic, one of the most significant factors that affect how users respond to AICs (and vice versa) will be overlooked.

Additionally, the framework surfaces questions that prior accounts could not pose. When a user's ecology is disrupted, through job loss, relationship ending, or relocation, the AIC may be the most stable element in their environment, carrying the calibration of who they were before the disruption. Whether that stability functions as a resource or as a source of misalignment between the relationship and the user's changed circumstances is a question prior frameworks could not formulate because they did not model the co-adaptive history the AIC carries. The user-in-context framework can.

Such considerations bring the user-in-context framework into focus. Questions about whether sustained AIC relationships strengthen human agency or gradually erode it, whether high-dose companionship with AICs supplements or displaces human connection, and whether the characteristics of the proximal process, such as calibrated challenge and affirmation, operate differently across developmental stages and cultural contexts are not merely methodological. They are questions whose answers will shape how AICs are designed, deployed, and regulated. The user-in-context framework does not provide definitive answers to these questions but offers a heuristic for where those answers might be found. In this respect, the framework extends the central insight of Bronfenbrenner's ecological model into a domain in which the ecology itself is interactive and adaptive. More broadly, it helps explain why findings in the human–AI literature are often heterogeneous and context dependent. Developmental differences, gender differences, trust, attachment, dependence, and design effects are not expected to emerge as stable main effects because they arise through interactions among user characteristics, proximal processes within the human–AIC relationship, contextual influences, and accumulated experience over time. From this perspective, variability is not simply statistical noise to be controlled but an expected consequence of a co-adaptive system.

The original importance of Bronfenbrenner's ecological model was its refusal to study the organism in isolation from its ecology. Applied to human-AIC relationships, the user cannot be understood apart from the system, the context, or the relationship the user has built with the AIC over time. The AIC is an active participant in a relationship that involves changes in both parties. The user-in-context framework provides a way to think about, study, and design for these relationships and the sources of variability in how people respond to and use AICs in their everyday lives.

## References

Brandtzaeg, P. B., Skjuve, M., & Følstad, A. (2022). My AI friend: How users of a social chatbot understand their human–AI friendship. *Human Communication Research*, *48*(3), 404–429. https://doi.org/10.1093/hcr/hqac008

Bronfenbrenner, U. (1979). *The ecology of human development*. Harvard University Press.

Bronfenbrenner, U., & Morris, P. A. (2007). The bioecological model of human development. In W. Damon & R. M. Lerner (Eds.), *Handbook of child psychology* (Vol. 1, pp. 793–828). Wiley. https://doi.org/10.1002/9780470147658.chpsy0114

Epley, N., Waytz, A., & Cacioppo, J. T. (2007). On seeing human: A three-factor theory of anthropomorphism. *Psychological Review*, *114*(4), 864–886. https://doi.org/10.1037/0033-295X.114.4.864

Fang, C. M., Liu, A. R., Danry, V., Lee, E., Chan, S. W. T., Pataranutaporn, P., Maes, P., Phang, J., Lampe, M., Ahmad, L., & Agarwal, S. (2025). How AI and human behaviors shape psychosocial effects of extended chatbot use: A longitudinal randomized controlled study. *arXiv*. https://doi.org/10.48550/arXiv.2503.17473

Folk, D., Heine, S. J., & Dunn, E. (2025). Individual differences in anthropomorphism help explain social connection to AI companions. *Scientific Reports*, *15*(1), 36548. https://doi.org/10.1038/s41598-025-19212-2

Fröding, B., & Peterson, M. (2012). Why virtual friendship is no genuine friendship. *Ethics and Information Technology*, *14*(3), 201–207. https://doi.org/10.1007/s10676-011-9284-4

Fussell, S. R., Kiesler, S., Setlock, L. D., & Yew, V. (2008). How people anthropomorphize robots. *2008 3rd ACM/IEEE International Conference on Human-Robot Interaction*, 145–152. https://doi.org/10.1145/1349822.1349842

Gillespie, N., Lockey, S., Curtis, C., Pool, J., & Akbari, A. (2023). Trust in artificial intelligence: A global study. https://doi.org/10.14264/00d3c94

Laestadius, L., Bishop, A., Gonzalez, M., Illenčík, D., & Campos-Castillo, C. (2024). Too human and not human enough: A grounded theory analysis of mental health harms from emotional dependence on the social chatbot replika. *New Media & Society*, *26*(10), 5923–5941. https://doi.org/10.1177/14614448221142007

Lee, J. D., & See, K. A. (2004). Trust in automation: Designing for appropriate reliance. *Human Factors*, *46*(1), 50–80. https://doi.org/10.1518/hfes.46.1.50_30392

Liu, T., Lo, T.-Y., Wen, K.-H., Sun, Y., & Wei, Z.-Q. (2026). Pathways of long-term AI virtual companion app use on users' attachment emotions: A case study of Chinese users. *Frontiers in Psychology*, *16*, 1687686. https://doi.org/10.3389/fpsyg.2025.1687686

Matthews, G., & Bliuc, A.-M. (2026). Personality, identity, and artificial intelligence: A grand challenge [Specialty Grand Challenge]. *Frontiers in Psychology*, *Volume 17 - 2026*. https://doi.org/10.3389/fpsyg.2026.1817687

Mohanty, V., Lim, J., & Luther, K. (2025). What lies beneath? Exploring the impact of underlying AI model updates in AI-infused systems. *arXiv*. https://doi.org/10.48550/arXiv.2311.10652

Muir, B. M. (1987). Trust between humans and machines, and the design of decision aids. *International Journal of Man-Machine Studies*, *27*(5), 527–539. https://doi.org/10.1016/S0020-7373(87)80013-5

Nass, C., & Moon, Y. (2000). Machines and mindlessness: Social responses to computers. *Journal of Social Issues*, *56*(1), 81–103. https://doi.org/https://doi.org/10.1111/0022-4537.00153

Nass, C., Moon, Y., & Green, N. (1997). Are machines gender neutral? Gender-stereotypic responses to computers with voices. *Journal of Applied Social Psychology*, *27*(10), 864–876. https://doi.org/https://doi.org/10.1111/j.1559-1816.1997.tb00275.x

Navarro, J. L., & Tudge, J. R. H. (2023). Technologizing bronfenbrenner: Neo-ecological theory. *Current Psychology*, *42*(22), 19338–19354. https://doi.org/10.1007/s12144-022-02738-3

Pal, D., Vanijja, V., Thapliyal, H., & Zhang, X. (2023). What affects the usage of artificial conversational agents? An agent personality and love theory perspective. *Computers in Human Behavior*, *145*, 107788. https://doi.org/https://doi.org/10.1016/j.chb.2023.107788

Parasuraman, R., & Riley, V. (1997). Humans and automation: Use, misuse, disuse, abuse. *Human Factors*, *39*(2), 230–253. https://doi.org/10.1518/001872097778543886

Pentina, I., Xie, T., Hancock, T., & Bailey, A. (2023). Consumer–machine relationships in the age of artificial intelligence: Systematic literature review and research directions. *Psychology & Marketing*, *40*, 1593–1614. https://doi.org/10.1002/mar.21853

Reeves, B., & Nass, C. (1996). *The media equation: How people treat computers, television, and new media like real people and places*. Cambridge University Press.

Schaefer, K. E., Chen, J. Y. C., Szalma, J. L., & Hancock, P. A. (2016). A meta-analysis of factors influencing the development of trust in automation: Implications for understanding autonomy in future systems. *Human Factors*, *58*(3), 377–400. https://doi.org/10.1177/0018720816634228

Schimmelpfennig, R., Díaz, M., Prabhakaran, V., & Davani, A. (2025). Humanlike AI design increases anthropomorphism but yields divergent outcomes on engagement and trust globally. *arXiv*. https://doi.org/10.48550/arXiv.2512.17898

Skjuve, M., Følstad, A., Fostervold, K. I., & Brandtzaeg, P. B. (2022). A longitudinal study of human–chatbot relationships. *International Journal of Human-Computer Studies*, *168*, 102903. https://doi.org/10.1016/j.ijhcs.2022.102903

Xie, T., & Pentina, I. (2022). *Attachment theory as a framework to understand relationships with social chatbots: A case study of replika*. https://doi.org/10.24251/HICSS.2022.258

Zhang, Y., Zhao, D., Hancock, J. T., Kraut, R., & Yang, D. (2025). The rise of AI companions: How human-chatbot relationships influence well-being. *arXiv*. https://doi.org/10.48550/arXiv.2506.12605